\documentclass[a4paper]{jpconf}
\usepackage{graphicx}

\begin{document}

\title{Method for finding the critical temperature of the island in a SET structure}

\author{J.~J.~Toppari$^1$, T.~K\"uhn$^1$, A.~P.~Halvari$^1$, G.~S.~Paraoanu$^2$}

\address{$^1$ NanoScience Center, Department of Physics, University of Jyv\"askyl\"a,
P.O.~Box 35 (YN), FIN-40014 University of Jyv\"askyl\"a, Finland}

\address{$^2$ Low Temperature Laboratory, Helsinki University of Technology,
P.O. Box 5100, FIN-02015 TKK, Finland}

\ead{paraoanu@cc.hut.fi}

\begin{abstract}
We present a method to measure the critical temperature of the island of a superconducting single electron transistor.
The method is based on a sharp change in the slope of the zero-bias conductance as a function of temperature.
We have used this method to determine the superconducting
phase transition temperature of the Nb island of an superconducting single electron transistor with Al leads.
We obtain $T_\mathrm{c}^\mathrm{Nb}$ as high as $8.5$ K and gap energies up to $\Delta_\mathrm{Nb}\simeq 1.45$ meV.
By looking at the zero bias conductance as a function of magnetic field instead of temperature, also the critical
field of the island can be determined.
Using the orthodox theory,
we have performed extensive numerical simulations of charge transport properties in the SET at temperatures comparable to the
gap, which match very well the data, therefore providing a solid theoretical basis for our method.


\end{abstract}

The single-electron transistor is a device with remarkable properties, which have been intensively studied
over the last almost two decades \cite{single1}. The first application of such structures have been in the area of electrometry
\cite{rfset}, but recently new developments have promoted the superconducting version of this transistor as one of the
main contenders for realizing the so-called charge quantum bits \cite{chargequbits}. These devices could become the
building blocks of future quantum computers, in architectures that allow exchange of quantum information between
them via transmission lines \cite{architectures}.

These new areas of
interest have created a demand for high performance
nanofabrication techniques and there has been a strong
motivation to develop a technique for a reliable fabrication of a small
niobium-based Josephson junction.
Niobium (Nb) would provide more reliable performance in many of these devices due
to its large superconducting gap $\Delta_\mathrm{Nb}\approx 1.5$ meV (in bulk) as compared to
aluminium, $\Delta_\mathrm{Al}\approx 0.2$ meV, which has been the
material of choice in nanofabrication for many years due to its easier processability. A larger superconducting
gap would provide a better suppression of the undesirable quasiparticle tunnelling
and also would yield larger Josephson coupling energy $E_\mathrm{J}$, in
ultrasmall junctions.

The conventional e-beam shadow evaporation (self-alignment)
technique, has been applied successfully for soft metals like Al, Cu and Pb.
Yet, it is known to be difficult to apply this method for refractory metals like Nb
due to the poor quality of the produced films, resulting in a  critical temperature $T_\mathrm{c}$ well below the bulk value.
To circumvent this problem, several fabrication methods have been proposed, and different groups have been able to obtain good quality small-capacity single
Nb-Al and Nb-Nb junctions \cite{single} and
Nb-based single electron transistors \cite{us}.

In all these devices, since there is no easy direct electrical access to the island,
the typical way to determine the critical temperature is to produce separate samples with films
of the same thickness, and measure the critical temperature of those. However, the critical temperature of the Nb island can be
quite different from that of a film, since the island's dimensions are small, comparable with the superconducting
coherence length ($\xi_\mathrm{Nb}=38$ nm in the bulk). Of course, determination of the gap
at low temperatures from tunneling data provides as well an indirect measure of the critical temperature via the BCS relation:
but again, for lower-dimensional superconducting objects the corrections to this law could be significant.

In this article we present a method to determine directly the critical temperature, and also the critical magnetic field, of a Nb island in superconducting
Al/AlOx/Nb/AlOx/Al-SETs fabricated with the standard self-alignment lithography
process using a slightly modified recipe.

The fabrication technology for our samples has been described elsewhere \cite{single,jussi}. The measurements were carried out in a small dilution refrigerator (Nanoway, PDR50)
with the base temperature below 100 mK and well-filtered electrical lines.

\begin{figure}[htb]
\includegraphics[width=140truemm]{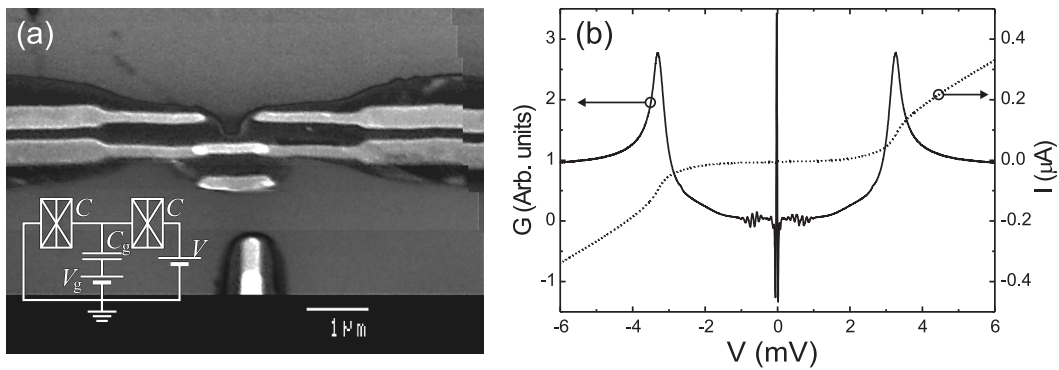}
\caption{ (a) SEM image of one of the measured samples with inset showing the schematics used for measuring the
SET. (b) The $I$-$V$ characteristics and $dI/dV$-curve measured from a sample with
$E_\mathrm{C}\approx 35$ $\mu$eV and $E_\mathrm{J}\approx 92$ $\mu$eV.
}
\label{one}
\end{figure}

In Fig. \ref{one}(a) we show an SEM picture of one of the
samples and its IV characteristic. In the SEM image, brighter lines
are niobium and darker lines are aluminium. Besides the existence of a quasiparticle voltage threshold and the Josepshon current,
the IV displays a number of features, such as Cooper-pair resonances and low-bias excess currents which have been analyzed elsewhere \cite{jussi}.

The charging energies of the samples were derived from the normal state conductance
curve measured at 4.2 K with magnetic field of $B \sim 5$ T. The obtained
charging energies varied between $E_\mathrm{C}= e^{2}/2C  \approx 34-78$ $\mu$eV. The
Josephson coupling energy was found from the  Ambegaogar-Baratoff formula for
two different superconductors, which yielded $E_\mathrm{J}\approx
41-123$ $\mu$eV.

In Fig.~\ref{one}(b) the $I$-$V$ characteristics and a $dI/dV$-curve measured at the temperature
$T \approx 200$ mK are shown from one of the samples with $E_\mathrm{C}\sim 35$ $\mu$eV and $E_\mathrm{J}\approx 92$ $\mu$eV.
The gap -- with a width of $4(\Delta_\mathrm{Al}+\Delta_\mathrm{Nb})/e$ -- is clearly
visible and the maxima in the $dI/dV$-curve yield $\Delta_\mathrm{Nb}\approx 1.45$ meV. Here we have assumed $\Delta_\mathrm{Al}\approx 0.2$ meV.
At bias voltages
below $V\approx 0.9$ mV we also see (Fig. \ref{one}(b), the $IV$ curve)
a series of gate-dependent, equally-spaced peaks, which appear only when the aluminium leads become superconducting.
They correspond to Cooper pair resonances occuring when charges are transported in the whole circuit \cite{jussi}.

\begin{figure}[htb]
\includegraphics[width=140truemm]{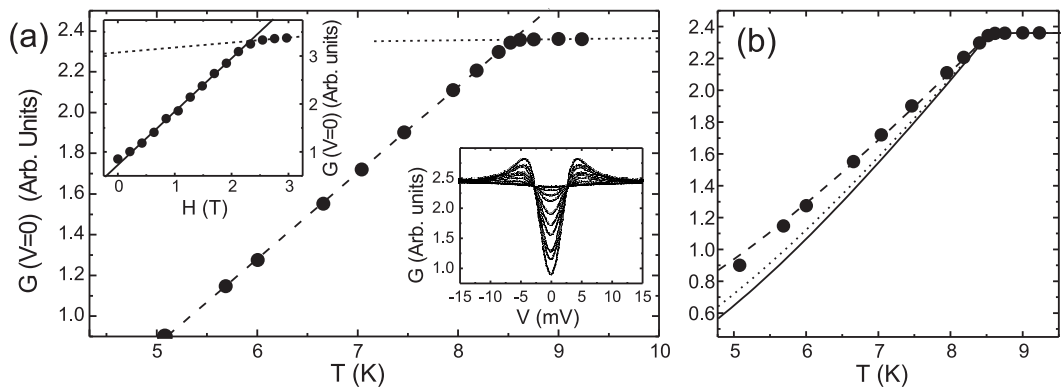}
\caption{ (a) Example of determining the critical temperature  $T_\mathrm{c}$
 in one of the samples.
Circles are the measured zero bias conductances obtained from the experimental
$dI/dV$-curves shown in the lower-right inset. The dashed line is a linear fit to the data below
$T_\mathrm{c}$. The slope is determined by the reduction of $\Delta_\mathrm{Nb}(T)$
as the temperature is increased. The dotted line is a linear fit to the data above $T_\mathrm{c}$
and is rather flat due to the much weaker temperature dependence of the Coulomb blockade.
In the left inset we present similar data for magnetic field instead of temperature,
a graph which we used for determining the upper critical field of Nb island.

(b) Data points (circles) and numerical calculations for charging energy, $E_\mathrm{C} =
35$ $\mu$eV. The continuous line corresponds to junctions without excess current, that is
 $\sigma = \Gamma = 0$; the dotted line was calculated with $\sigma = 0.38$ meV, $\Gamma = 37.5$ $\mu$eV,
and a best fit around the critical temperature was obtained (dashed line) for $\sigma = 0.38$ meV, $\Gamma =161.25$ $\mu$eV.}

\label{figure2}
\end{figure}

Experimentally, our method consists of determining the critical temperature $T_\mathrm{c}^\mathrm{Nb}$
as well the upper critical field $H_{\mathrm{c},2}^\mathrm{Nb}$ from the zero bias conductance of the sample as a function of temperature
or magnetic field, respectively. An example is shown
in Fig.~\ref{figure2} where the sudden transition from the strong temperature dependence
determined by the reduction in $\Delta_\mathrm{Nb}(T)$, to a much weaker temperature dependence
(due only to Coulomb blockade) is clearly visible at the phase transition. The critical
temperatures obtained were 7.8, 8.1 and 8.5 K in the three best samples and above 7.5 K
in all others. This is already very close to $T_\mathrm{c}^\mathrm{Nb}\approx 9$ K
in bulk Nb. The obtained critical fields were between $H_{\mathrm{c},2}^\mathrm{Nb}\approx 2.5 - 4.5$ T.
In this case the transition to the normal state happens gradually between the thermal critical field $H_\mathrm{c,1}^\mathrm{Nb}$
and the upper critical field $H_{\mathrm{c},2}^\mathrm{Nb}$ where superconductivity is completely suppressed
(thin, disordered Nb films are type-II superconductors),
as seen on the left inset of Fig.~\ref{figure2}(a).

To model or data, we have used the standard orthodox theory \cite{single1}
for charge transport in NISIN single-electron transistors (at the temperatures of interest, around the critical temperature
of Nb, the Al leads are normal). We calculate numerically the tunneling probabilities in each junction for a given superconducting gap,
and to find the currents and conductances at a given bias voltage we solve the corresponding master equation for charge transport across the structure for each given temperature. From the low-bias Cooper
pair resonance data (which depend on the charging energy) we extract, as in \cite{jussi},  $E_\mathrm{C} =
35$ $\mu$eV. For the gap, which is used to calculate the tunneling probabilities, we use the standard BCS gap equation at finite temperature
\begin{eqnarray*}
\frac{1}{N(0)V}&=& \int\limits_{0}^{\infty}d\epsilon\frac{\tanh(\beta\sqrt{\epsilon^2+\Delta(T)^2}/2)}
                                                 {\sqrt{\epsilon^2+\Delta(T)^2}}\,,
\end{eqnarray*}
where $N(0)$ is the density of states at the Fermi level, $V$ is the BCS interaction, $\beta = 1/k_{B}T$ and the Debye energy is taken much larger
than the gap.
Since $\Delta(T=T_C)=0$, we can write
\begin{eqnarray*}
0 &=& \int\limits_{0}^{\infty}d\epsilon\left\{\frac{\tanh(\beta\sqrt{\epsilon^2+\Delta(T)^2}/2)}
{\sqrt{\epsilon^2+\Delta(T)^2}}
-\frac{\tanh(\beta_C\epsilon /2)}{\epsilon}\right\}\,,
\end{eqnarray*}
an integral equation which is used to determine numerically the gap value at a given temperature T.

Also, when calculating the tunneling probabilities, in order to take into account the excess subgap currents present in Nb-based junctions,
we have used
a life-time broadening $\Gamma$ of the quasiparticle energies, resulting in a
density of states \cite{jussi}
\begin{equation}
\rho (E) = \int_{0}^{\infty} d\Delta_\mathrm{Nb} {\cal P}(\Delta_\mathrm{Nb})\left\vert {\rm Re} \left( \frac{E - i \Gamma}{\sqrt{(E- i \Gamma )^2 - \Delta_{\rm 
Nb}^2}} \right) \right\vert .
\end{equation}
where ${\cal P}$ is the probability density associated with a certain value of the gap, where the
function ${\cal P}$ is a Gaussian of standard distribution $\sigma$. The value of $\sigma$ was determined from
this sample using the low-bias $IV$ plots, as described in \cite{jussi}: its value is set by the fabrication process and does
not depend on temperature.
The life-time broadening $\Gamma$ depends in general on temperature: its value at $T = 95.7$ mK was detemined in \cite{jussi} as $\Gamma = 37.5$ $\mu$eV.
We assume that this dependence is mild near the phase transition so that $\Gamma$ can be considered constant: we find,
as a best fit to the data, $\Gamma =161.25$ $\mu$eV.
We note also that at 5-6 K this value does not give a good fit, and a better match for the data
would require indeed a lower value of $\Gamma$, corresponding to a lower temperature.
Fig. \ref{figure2}(b) summarizes these findings.

We described a new method for determining the critical temperature of a Nb island which
forms the active element in a superconducting single electron transistor. The method is based on the experimental observation that,
as the system is cycled through the phase transition, the slope of the zero-bias conductance as a function of temperature changes abruptly
from a finite value (in the superconducting state) to almost zero (in the normal state).

\ack

This work was supported by the Academy of Finland (Projects No. 00857, No. 7111994, and No. 7118122).


\section*{References}


\begin{thebibliography}{99}

\bibitem{single1} Grabert H, Devoret M H (eds.) 1992, {\it Single Charge Tunneling} (Plenum, New York).

\bibitem{rfset} Schoelkopf R J {\it et. al.} 1998, Science {\bf 280}, 1238; Devoret M H, Schoelkopf R J 2000,
Nature {\bf 406}, 1039

\bibitem{chargequbits} Nakamura Y, Pashkin Y A, Tsai J S,
 Nature 1999, {\bf 398}, 786; Vion D {\it at. al} 2002, Science {\bf 296}, 886; Pashkin Yu. A. {\it et. al.} 2003, 
Nature {\bf 421}, 823; Yamamoto T {\it et. al.} 2003, Nature {\bf 425}, 941; ; Duty T, Gunnarsson D,
Bladh K, Delsing P 2004, Phys. Rev. B {\bf 69}, 140503(R).

\bibitem{architectures} Blais A 2004 {\it et. al} Phys. Rev. A {\bf 69}, 062320; Wallraff A {\it et. al.} 2004, Nature (London)
{\bf 431}, 162;
Blais A {\it et. al.} 2007,
Phys. Rev. A {\bf 75}, 032329;
Paraoanu G S 2006, Phys. Rev. B {\bf 74}, 140504 (R);
Li J, Chalapat K, Paraoanu G S 2008 Entanglement of superconducting qubits via microwave fields: classical and quantum regimes
{\it Preprint}, arXiv:0803.0397

\bibitem{single} Paraoanu G S, Halvari A M 2003, Rev. Adv. Mater. Sci. {\bf 5}, 292

\bibitem{us} Zorin A B, Lotkhov S V, Zangerle H, and Niemeyer J, 2000 J. Appl. Phys. {\bf 88}, 2665; Dolata R, Scherer H, Zorin A B,
Krupenin V A, Niemyer J 2002, Appl. Phys. Lett. {\bf 80} 2776;
Kim N {\it et. al.} 2003, Physica B {\bf 329-333}, 1519; Watanabe M, Nakamura Y, Tsai J-S, Appl. Phys. Lett. {\bf 84} 410;
Savin A M {\it et. al.} 2007 Appl. Phys. Lett. {\bf 91}, 063512

\bibitem{jussi} Toppari J J {\it et. al} 2007, Phys. Rev. B {\bf  76}, 172505



\end{thebibliography}
\end{document}